\documentclass[preprint,showpacs,amsmath,amssymb]{revtex4}
\usepackage{dcolumn}
\usepackage{bm}

\makeatletter

\def\compoundrel#1\over#2{\mathpalette\compoundreL{{#1}\over{#2}}}
\def\compoundreL#1#2{\compoundREL#1#2}
\def\compoundREL#1#2\over#3{\mathrel
  {\vcenter{\hbox{$\m@th\buildrel{#1#2}\over{#1#3}$}}}}
\makeatother
\newcommand{\be}{\begin{equation}}
\newcommand{\ee}{\end{equation}}
\newcommand{\bea}{\begin{eqnarray}}
\newcommand{\eea}{\end{eqnarray}}
\newcommand{\bref}[1]{(\ref{#1})}
\newcommand{\mapright}[1]{%
	\smash{\mathop{%
	\hbox to 1cm{\rightarrowfill}}\limits^{#1}}}

\begin{document}
\draft
\title{{Lepton Mixing Matrix Element $U_{13}$ and New Assignments of \\ 
Universal Texture for Quark and Lepton  Mass Matrices}}

\author{Koichi MATSUDA}
\address{
Graduate school of Science, 
Osaka University, Toyonaka, Osaka 560-0043, Japan}
\author{Hiroyuki NISHIURA}
\address{
Department of General Education, 
Junior College of Osaka Institute of Technology, \\
Asahi-ku, Osaka 535-8585, Japan}

\date{February 29, 2004}

\begin{abstract}

We reanalyze the mass matrix model of quarks and leptons that gives a unified description of 
quark and lepton mass matrices with the same texture form. 
By investigating possible types of assignment for the texture 
components of the lepton mass matrix, we find that a different assignment 
for neutrinos than for charged leptons can also lead to consistent values of 
the MNSP lepton mixing matrix. We also find that the predicted value for the lepton mixing matrix 
element $U_{13}$ of the model depends on the assignment. 
A proper assignment will be discriminated by future 
experimental data for $U_{13}$. 
\end{abstract}


\maketitle


Recent neutrino oscillation experiments\cite{skamioka} have highly
suggested a nearly bi-maximal lepton mixing $(\sin^2 2\theta_{12}\sim 1$, 
$\sin^2 2\theta_{23}\simeq 1)$ in contrast with the small quark mixing.
In order to reproduce these lepton and quark mixings, 
mass matrices of various structures with texture zeros have been investigated 
in the literature\cite{fritzsch}-\cite{Ramond}.
Recently a mass matrix model based on a discrete symmetry $Z_3$ and 
a flavor $2 \leftrightarrow 3$ symmetry has been proposed\cite{Koide} 
with following universal structure for all quarks and leptons 
\begin{equation}
P^\dagger\left(
\begin{array}{lll}
\ 0 & \ A & \ A \\
\ A & \ B & \ C \\
\ A & \ C & \ B \\
\end{array}
\right)P^\dagger, \ \ 
\end{equation}
where $P$ is a diagonal phase factor.
It was pointed out\cite{Koide} that this structure leads to reasonable values of 
the Cabibbo--Kobayashi--Maskawa (CKM) \cite{CKM} quark mixing as well as lepton mixing. 
An assignment (we refer it as Type A) for the texture components with 
$A=\pm\sqrt{\frac{m_2m_1}{2}}$, $B=\frac{1}{2} \left(m_3+m_2-m_1\right)$, and 
$C=-\frac{1}{2}\left(m_3-m_2+m_1\right)$ with the i-th genaration mass $m_i$ has been proposed 
both for up and down quarks in Ref.\cite{Koide}. 
Unfortunately, the type A assignment leads to a somewhat small predicted value 
for the CKM quark mixing matrix element $|V_{ub}|=0.0020 - 0.0029$  
with respect to the present experimental value $|V_{ub}|=0.0036  \pm 0.0007$\cite{PDG}.
Subsequently, it has been pointed out\cite{Matsuda} that there exist other possible 
new assignments for texture components 
and that the combination of the new assignment(we refer it Type B) with 
$A=\pm\sqrt{\frac{m_3m_1}{2}}$, $B=\frac{1}{2} \left(m_3+m_2-m_1\right)$, and 
$C=\frac{1}{2}\left(m_3-m_2-m_1\right)$ for up quarks, 
and the previously proposed type A assignment for down quarks is phenomenologically 
favorable for reproducing the consistent values of CKM quark mixing. 
By taking the type A assignment both for neutrinos and charged leptons, 
the authors in Ref. \cite{Koide} have obtained the consistent values for the lepton mixing matrix of the model  
with the present experimental data . 
In the present paper, stimulated by the success in the quark sector, 
we discuss the neutrino masses and lepton mixings of the model with use of the new assignments 
for charged leptons and neutrions. 
\par
Our mass matrices 
\(M_u\), \(M_d\), \(M_\nu\), and  \(M_e\) 
for up quarks (\(u,c,t\)), down quarks (\(d,s,b\)), 
neutrinos (\(\nu_e,\nu_\mu,\nu_\tau\)) and 
charged leptons (\(e,\mu,\tau\)), 
respectively are given as follows\cite{Koide}:
\begin{equation}
M_f = P_{f}^\dagger\widehat{M}_fP_{f}^\dagger, \
\quad \quad \quad \quad 
\end{equation}
with
\begin{equation}
\widehat{M}_f=
\left(
\begin{array}{lll}
\ 0 & \ A_f & \ A_f \\
\ A_f & \ B_f & \ C_f \\
\ A_f & \ C_f & \ B_f \\
\end{array}
\right) \ \, \left(f=u,d,\nu,e\right),
\label{texture}
\end{equation}
where $P_{f}$ is the diagonal phase matrix 
and $A_f$, $B_f$, and $C_f$ are real parameters.
\par  
Hereafter, for brevity, we will omit the flavor index.
The eigenmasses of Eq. \bref{texture} are given by
$
\frac{1}{2}\left[B+C-\sqrt{8A^2 + (B+C)^2}\right]$ ,
$\frac{1}{2}\left[B+C+\sqrt{8A^2 + (B+C)^2}\right]$ , and
$B-C$. 
Therefore, there are three types of assignments for the texture 
components of \(\widehat{M}\) according to the assignments for the eigenmass $m_i$.
\par
(i)\ Type A: 
\begin{eqnarray}
-m_1& =&
\frac{1}{2}
\left[B+C-\sqrt{8A^2 + (B+C)^2}
\right] ,\\
m_2& =&\frac{1}{2}
\left[B+C+\sqrt{8A^2 + (B+C)^2}
\right] ,\\
m_3& =&B-C. 
\end{eqnarray}
This is the case for which $B-C$ has the largest value. In this type, 
the texture components of \(\widehat{M}\) are expressed 
in terms of $m_i$ as 
\begin{eqnarray}
A  =\pm\sqrt{\frac{m_2m_1}{2}}  , \quad
B  =\frac{1}{2} \left(m_3+m_2-m_1\right) ,\quad 
C  =-\frac{1}{2}\left(m_3-m_2+m_1\right) .\label{eq2003} 
\end{eqnarray}
The orthogonal matrix \(O\) that diagonalizes 
\(\widehat{M}\) [$(O^{T}\widehat{M}O=diag(-m_1,m_2,m_3)$] is 
given by 
\begin{equation}
O\equiv
\left(
\begin{array}{ccc}
{  c}  & { \pm s}& {0} \\
{\mp \frac{s}{\sqrt{2}}} & {\frac{c}{\sqrt{2}}} & {-\frac{1}{\sqrt{2}}} \\
{\mp \frac{s}{\sqrt{2}}} & {\frac{c}{\sqrt{2}}} & {\frac{1}{\sqrt{2}}}
\end{array}
\right). \label{eq990114} 
\end{equation}
Here \(c\) and \(s\) are defined by $c=\sqrt{\frac{m_2}{m_2+m_1}}$ and $s=\sqrt{\frac{m_1}{m_2+m_1}}$.
Note that the elements of \(O\) are independent of \(m_3\).  
This type A assignment is proposed in Ref.\cite{Koide}. 
\par
(ii)\ Type B: This assignment is obtained by exchanging $m_2$ and $m_3$ in type A:
\begin{eqnarray}
-m_1& =&
\frac{1}{2}
\left[B+C-\sqrt{8A^2 + (B+C)^2}
\right] ,\\
m_2& =&B-C,\\
m_3& =&\frac{1}{2}
\left[B+C+\sqrt{8A^2 + (B+C)^2}
\right]. 
\end{eqnarray}
In this type, the texture components of \(\widehat{M}\) are expressed as 
\begin{eqnarray}
A  =\pm\sqrt{\frac{m_3m_1}{2}}  ,\quad
B  =\frac{1}{2}\left(m_3 +m_2-m_1\right) ,\quad 
C  =\frac{1}{2}\left(m_3-m_2-m_1\right) .\label{eq2004} 
\end{eqnarray}
The orthogonal matrix \(O^\prime\) that diagonalizes 
\(\widehat{M}\) [$(O^{\prime T}\widehat{M}O^\prime=diag(-m_1,m_2,m_3)$] is 
given by 
\begin{equation}
O^\prime\equiv
\left(
\begin{array}{ccc}
{ c^\prime} & {0} & {\pm s^\prime} \\
{\mp \frac{s^\prime}{\sqrt{2}}} & {\frac{1}{\sqrt{2}}} & {\frac{c^\prime}{\sqrt{2}}} \\
{\mp \frac{s^\prime}{\sqrt{2}}} & {-\frac{1}{\sqrt{2}}} & {\frac{c^\prime}{\sqrt{2}}}
\end{array}
\right). \label{eq990114B} 
\end{equation}
Here \(c^\prime\) and \(s^\prime\) are defined by $c^\prime=\sqrt{\frac{m_3}{m_3+m_1}}$ 
and $s^\prime=\sqrt{\frac{m_1}{m_3+m_1}}$.
\par
(iii)\ Type C: This assignment is obtained by exchanging $m_1$ for $m_2$ in type B. 
However, This type is not useful in the following discussions.
\par
The type B and type C are new assignments proposed in Ref.\cite{Matsuda}. 
Taking the type B assignment for up quarks and the type A for down quarks,
we have obtained consistent values for the CKM quark mixing matrix in Ref.\cite{Matsuda}. 
Stimulated by the above new assignment, Koide\cite{Koide2} has proposed 
a new universal texture of quark and lepton mass matrices which is an extension 
of our model with an extended flavor $2 \leftrightarrow 3$ symmetry.
\par
Now let us discuss the lepton mixing matrix of the model. 
In our model, \(M_{\nu}\) and \(M_e\) have the same zero texture with same or different assignments as follows:
\begin{eqnarray}
M_{\nu}&=&
P_{\nu}^\dagger\left(
\begin{array}{lll}
\ 0 & \ A_{\nu} & \ A_{\nu} \\
\ A_{\nu} & \ B_{\nu} & \ C_{\nu} \\
\ A_{\nu} & \ C_{\nu} & \ B_{\nu} \\
\end{array}
\right)P_{\nu}^\dagger \ , \\
M_{e}&=&
P_e ^\dagger\left(
\begin{array}{lll}
\ 0 & \ A_e & \ A_e \\
\ A_e & \ B_e & \ C_e \\
\ A_e & \ C_e & \ B_e \\
\end{array}
\right)P_e^\dagger \label{eq5} \ ,
\end{eqnarray}
where \(P_{\nu}\) and \(P_e\) are the $CP$-violating phase factors.
We find that the assignments that are consistent with the present experimental data are following two cases:
\par 
Case(i): Type A assignment both for neutrinos and charged leptons are taken. 
\par
Case(ii): Type B assignment for neutrinos and type A for charged leptons are taken.  

The other possible cases fail to reproduce consistent lepton mixing. 
Since the case(i) was discussed in Ref.\cite{Koide}, let us discuss the case(ii) in this paper. 
In case(ii), $M_f = P_{f}^\dagger\widehat{M}_fP_{f}^\dagger  \ \ (f=\nu,e)$ 
are diagonalized by the biunitary transformation as
$D_f  =  U_{Lf}^\dagger M_f U_{Rf}$ , 
where $U_{L\nu}\equiv P_{\nu}^\dagger O_{\nu}$, $U_{R\nu}\equiv P_{\nu} O_{\nu}$, 
$U_{Le}\equiv P_e^\dagger O_e^\prime$, and $U_{Re}\equiv P_e O_e^\prime$. 
Here $O_\nu$ and $O_e^\prime$ are orthogonal matrices 
that diagonalize $\widehat{M}_\nu$ and $\widehat{M}_e$, respectively. 
Thus, we obtain the Maki--Nakagawa--Sakata--Pontecorv (MNSP) \cite{MNSP} lepton mixing 
matrix \(U\) as follows:
\begin{equation}
U=U^\dagger_{Le}U_{L\nu}=O^{\prime T}_{e}P_{e}P^\dagger_{\nu} O_{\nu}
=\left(
\begin{array}{ccc}
c^\prime_e c_\nu+\rho_\nu s^\prime_e s_\nu \quad & c^\prime_e s_\nu-\rho_\nu s^\prime_e c_\nu 
\quad & -{\sigma_\nu}s^\prime_e \\
-{\sigma_\nu}s_\nu \quad & {\sigma_\nu}c_\nu \quad & \rho_\nu \\
s^\prime_e c_\nu-{\rho_\nu}c^\prime_e s_\nu \quad & s^\prime_e s_\nu+{\rho_\nu}c^\prime_e c_\nu 
\quad & {\sigma_\nu}c^\prime_e \\
\end{array}
\right),\label{MNS} 
\end{equation}
where  \(\rho_\nu\) and \(\sigma_\nu\) are defined by 
\begin{equation}
\rho_\nu=\frac{1}{2}(e^{i\delta_{\nu3}}+e^{i\delta_{\nu2}})
=\cos\frac{\delta_{\nu3} - \delta_{\nu2}}{2} \exp\left[ i
\left( \frac{\delta_{\nu3} + \delta_{\nu2}}{2} \right)\right] \ ,
\end{equation}
\begin{equation}
\sigma_\nu=\frac{1}{2}(e^{i\delta_{\nu3}}-e^{i\delta_{\nu2}})
= \sin\frac{\delta_{\nu3} - \delta_{\nu2}}{2} 
\exp\left[ i \left( \frac{\delta_{\nu3} + \delta_{\nu2}}{2}+ \frac{\pi}{2}
\right)\right] \ . 
\end{equation}
Here we have set \(P \equiv P_{e}P^\dagger_{\nu} \equiv 
\mbox{diag}(e^{i\delta_{\nu1}}, e^{i\delta_{\nu2}},e^{i\delta_{\nu3}})\), 
and we have taken \(\delta_{\nu1}=0\) without 
loss of generality.
Then, the explicit magnitudes of the components of \(U\) are expressed as
\begin{align}
\left| U_{11}\right|
& \simeq \sqrt{\frac{m_2}{m_2+m_1}}, &
\left| U_{12}\right| & \simeq \sqrt{\frac{m_1}{m_2+m_1}}, &
\left|U_{13}\right| & = \sqrt{\frac{m_e}{m_\tau+m_e}}\sin\frac{\delta_{\nu3} - \delta_{\nu2}}{2}, \\ 
\left|U_{22}\right| & = \sqrt{\frac{m_2}{m_2+m_1}}\sin\frac{\delta_{\nu3} - \delta_{\nu2}}{2}, &
\left|U_{23}\right| & = \cos\frac{\delta_{\nu3}-\delta_{\nu2}}{2},&
\left|U_{33}\right| &\simeq \sin\frac{\delta_{\nu3}-\delta_{\nu2}}{2}. 
\end{align}
Therefore, we obtain 
\begin{eqnarray}
\tan^2\theta_{\mbox{{\tiny solar}}}& =&\frac{|U_{12}|^2}{|U_{11}|^2}\simeq \frac{m_1}{m_2}\ ,\label{eq30300}\\
\sin^2 2\theta_{\mbox{{\tiny atm}}}& =&4|U_{23}|^2|U_{33}|^2\simeq \sin^2(\delta_{\nu3} - \delta_{\nu2})\ ,
\label{eq30310} \\
|U_{13}|^2 &\simeq& \frac{m_e}{m_\tau}\sin^2\frac{\delta_{\nu3} - \delta_{\nu2}}{2} .\label{eq30320}
\end{eqnarray}
Note that $|U_{ij}|$ are almost independent of $(\delta_{\nu3} + \delta_{\nu2})$.  
Therefore, the independent parameters in $|U_{ij}|$ are 
$\theta^\prime_e = \tan^{-1}(m_e/m_\tau)$,
$\theta_\nu = \tan^{-1} (m_1/m_2)$, and ($\delta_{\nu3}-\delta_{\nu2}$). 
Since $\theta^\prime_e$ is already fixed by the charged lepton masses,  
$\theta_\nu$ and $(\delta_{\nu3} -\delta_{\nu2})$ are adjustable parameters in our model.
\par
Let us estimate the values of $\theta_\nu$ and  $(\delta_{\nu 3}-\delta_{\nu 2})$ by fitting the experimental data.
In the following discussions we consider the normal mass hierarchy 
\(\Delta m_{23}^2=m_3^2-m_2^2>0\) for the neutrino mass. 
In this situation, we can ignore the evolution effects.
They only give small correction effects enough to be neglected. 
Generally speaking, we must consider the renormalization group equation (RGE) effect 
when the neutrino masses
have the inverse hierarchy or almost degenerate ($m_1 \simeq m_2$). 
However, these scenarios are denied from (21) and (26).
Hence, the neutrino masses must have the normal mass hierarchy
and we do not need to consider the RGE effects. 
\par
When we use the expressions (19)-(23) at the weak scale, the parameters $\delta_{\nu2}$ and $\delta_{\nu3}$ 
do not mean the phases that are evolved from those at the unification scale. 
Hereafter, we use the parameters $\delta_{\nu2}$ and $\delta_{\nu3}$ as phenomenological parameters 
that approximately satisfy the relations (19)-(23) at the weak scale. 
 
We have\cite{Garcia}
\begin{equation}
|U_{13}|_{\mbox{\tiny exp}}^2 <  0.054 \ \label{mat20820} \ ,
\end{equation}
from the CHOOZ\cite{chooz}, solar\cite{sno}, and atmospheric neutrino 
experiments\cite{skamioka} with $3\sigma$ range. 
From the global analysis of the SNO solar neutrino experiment\cite{sno,Garcia} with $3\sigma$ range, we have 
\begin{eqnarray} 
\Delta m_{12}^2=m_2^2-m_1^2= \Delta m_{\mbox{{\tiny sol}}}^2
=(5.2-9.8) \times 10^{-5}\, \mbox{eV}^2, \label{mat20830} \\
\tan^2 \theta_{12}=\tan^2 \theta_{\mbox{{\tiny sol}}}=0.29-0.64 \ , 
\quad \quad \quad \quad
\label{eq20501}
\end{eqnarray}
for the large mixing angle (LMA) MSW solution.
From the atmospheric neutrino experiment\cite{skamioka,Garcia}, 
we also have
\begin{eqnarray}
\Delta m_{23}^2=m_3^2-m_2^2 \simeq \Delta m_{\mbox{{\tiny atm}}}^2
= (1.4-3.4) \times 10^{-3}\, \mbox{eV}^2, \label{mat20831}\\ 
\tan^2 \theta_{23} \simeq \tan^2 \theta_{\mbox{{\tiny atm}}}=0.49-2.2 \ , 
\quad \quad \quad \quad
\label{mat208302}
\end{eqnarray}
with $3\sigma$ range.
The observed fact $\tan^2 \theta_{\mbox{{\tiny atm}}} 
\simeq 1.0$ highly suggests
$\delta_{\nu3}-\delta_{\nu2}\simeq \pi/2$. 
Hereafter, for simplicity, we take
\begin{equation}
\delta_{\nu3}-\delta_{\nu2}=\frac{\pi}{2} \  .
\end{equation}
Under this constraint, the model predicts 
\begin{equation}
|U_{13}|^2={\frac{1}{2}}{\frac{m_e}{m_\tau+m_e}}=0.00014 \ ,
\ \mbox{or}~~\mbox{sin}^22\theta_{13}=0.00055 \ . \label{mat20870} 
\end{equation}
Here we have used the running charged lepton mass at \(\mu=\Lambda_X\)
\cite{Fusaoka}: $m_e(\Lambda_X)=0.325\ \mbox{MeV}$, 
$m_\mu(\Lambda_X)=68.6\ \mbox{MeV}$, 
and $m_\mu(\Lambda_X)=1171.4 \pm 0.2\ \mbox{MeV}$.
The value in Eq.(\ref{mat20870}) is consistent with the present experimental constraints 
Eq.(\ref{mat20820}) and can be checked in neutrino factories in future
\cite{cervera} which have sensitivity for sin$^22\theta_{13}$ as
$\mbox{sin}^22\theta_{13}\ge 10^{-5}$.
\par
From the mixing angle \(\theta_{\mbox{{\tiny sol}}}\) in the present model,  
we obtain
\begin{equation}
\frac{m_1}{m_2} \simeq \tan^2\theta_{\mbox{{\tiny sol}}}=0.29 - 0.64.
\label{ratio}
\end{equation}
Then, we obtain the neutrino masses
\begin{eqnarray}
m_1^2 & = & (0.48-6.8) \times 10^{-5} \  {\rm eV^2} \ ,\nonumber \\
m_2^2 & = & (5.7-16.6) \times 10^{-5} \  {\rm eV^2} \ , \\
m_3^2 & = & (1.4-3.4) \times 10^{-3} \  {\rm eV^2} \ .\nonumber
\end{eqnarray}
\par
Next let us discuss the CP-violation phases in the lepton mixing matrix.
The Majorana neutrino fields do not have the freedom of rephasing 
invariance, so that we can use only the rephasing freedom of $M_e$ 
to transform Eq.(\ref{MNS}) to the standard form
\begin{eqnarray}
& &U_{\rm std} 
= \mbox{diag}(e^{i\alpha_1^e},e^{i\alpha_2^e},e^{i\alpha_2^e}) 
\ U \ \mbox{diag}(e^{\pm i\pi/2},1,1)  \nonumber \\ 
& &= \left(
\begin{array}{ccc}
c_{\nu13}c_{\nu12} & c_{\nu13}s_{\nu12}e^{i\beta} & 
s_{\nu13}e^{i(\gamma-\delta_{\nu})} \\
(-c_{\nu23}s_{\nu12}-s_{\nu23}c_{\nu23}s_{\nu13} e^{i\delta_{\nu}})e^{-i\beta}
&c_{\nu23}c_{\nu12}-s_{\nu23}s_{\nu12}s_{\nu13} e^{i\delta_{\nu}} 
&s_{\nu23}c_{\nu13}e^{i(\gamma-\beta)} \\
(s_{\nu23}s_{\nu12}-c_{\nu23}c_{\nu12}s_{\nu13} e^{i\delta_{\nu}})e^{-i\gamma}
 & (-s_{\nu23}c_{\nu12}-c_{\nu23}s_{\nu12}s_{\nu13} 
e^{i\delta_{\nu}})e^{-i(\gamma-\beta)} 
& c_{\nu23}c_{\nu13}\\ 
\end{array}
\right) \ .
\label{majorana}
\end{eqnarray}
Here, \(\alpha_i^e\) comes from the rephasing in the charged lepton fields 
to make the choice of phase convention, and 
the specific phase \(\zeta\equiv{\pm \pi/2}\) is added on the right-hand side of \(U\) 
in order to change the neutrino eigenmass \(m_1\) to a positive quantity.
The CP-violating phase \(\delta_{\nu}\), the additional Majorana 
phase factors $\beta$ and $\gamma$ \cite{bilenky,Doi} 
in the representation Eq.(\ref{majorana}) are calculable and obtained as
\begin{eqnarray}
\delta_\nu  &=& \mbox{arg}
          \left[
             \frac{U_{12}U_{22}^*}{U_{13}U_{23}^*} + 
             \frac{|U_{12}|^2}{1-|U_{13}|^2}
          \right] 
 \simeq \mbox{arg} \left(
          \frac{U_{12} U_{22}^*}{U_{13} U_{23}^*} \right) \nonumber \\
  &\simeq& \frac{\pi}{2} \pm \frac{\pi}{2} -\frac{\delta_{\nu 3}+\delta_{\nu2}}{2}, \
 (\mbox{for } A_e=\pm|A_e|) \nonumber  \\
\beta & =& \mbox{arg} \left( \frac{U_{12}}{U_{11}}\right) + \zeta
  \simeq \frac{\pi}{2}\mp\frac{\pi}{2} +\zeta, \ (\mbox{for } A_\nu=\pm|A_\nu|) \nonumber  \\ 
\gamma  & =& \mbox{arg} \left( \frac{U_{13}}{U_{11}}\right) + \zeta
  \simeq - \frac{\pi}{2} +\zeta \ ,
\end{eqnarray}
by using the relation \(m_e \ll m_\tau \).
Note that the sign ambiguity in $A_e$ and $A_\nu$ can be absorbed into the phases $\delta_{\nu 2}$ 
and $\delta_{\nu 3}$ by the simultaneous redefinition of them. Hence, we
can also predict the averaged neutrino mass 
$\langle m_\nu \rangle$ which appears in the neutrinoless 
double beta decay\cite{Doi} as follows:
\begin{eqnarray}
\langle m_\nu \rangle & \equiv & \left| -m_1 U_{11}^2 +m_2 U_{12}^2
+m_3 U_{13}^2 \right| \nonumber \\
 & \simeq & \left|\frac{2m_1m_2}{m_1+m_2}-m_3\frac{m_e}{2m_\tau}
            e^{i(\delta_{\nu 3}+\delta_{\nu 2})} \right| \nonumber \\
&\simeq&  \frac{2m_1m_2}{m_1+m_2} \simeq 0.0017-0.0050\  {\rm eV} \ .
\end{eqnarray}
\par
Let us compare the result of the case (ii) with the case (i). In case (i), we have\cite{Koide}
\begin{equation}
|U_{13}|^2 \simeq \frac{m_e}{m_\mu}\sin^2\frac{\delta_{\nu3} - \delta_{\nu2}}{2} ,\label{eq40100}
\end{equation}
while the result of $\tan^2\theta_{\mbox{{\tiny solar}}}$ and $\sin^2 2\theta_{\mbox{{\tiny atm}}}$ are 
the same as Eqs. (\ref{eq30300}) and (\ref{eq30310}), respectively. 
Consequently, the values of $\theta_\nu$ and  $(\delta_{\nu 3}-\delta_{\nu 2})$ and the prediction for 
the neutrino masses of the model are also the same both for the cases (i)
and (ii). 
However, an interesting difference between the cases (i) and (ii) 
appears in the prediction for $|U_{13}|^2$. Namely 
the value of $|U_{13}|^2$ in the case (ii) are much smaller than in the case (i) as seen in 
Eqs.(\ref{eq30320}) and (\ref{eq40100}).  
Therefore, the cases (i) and (ii) are distinguishable by the experimental determination of 
$|U_{13}|^2$ in future, although they are both allowed from the present experimental data.
\par
In conclusion, we have reanalyzed the lepton mixing matrix 
using the mass matrix model of Ref.\cite{Koide} with the universal texture form. 
We have shown that the case (ii) with the type A assignment for $\widehat{M}_e$ and the type B for $\widehat{M}_\nu$ 
can also lead to consistent values for the lepton mixing matrix. 
The neutrino masses, CP-violating phases, and the averaged neutrino mass $\langle m_\nu \rangle$ 
in the case (ii) are predicted. 
It is also shown that the predicted value for $|U_{13}|$ depends 
on the assignment for the texture components of the mass matrices. 
Therefore, a proper assignment will be discriminated by future experimental data for $|U_{13}|$.

\vspace{5mm}
This work of K.M. was supported by the JSPS, No. 3700.



\end{document}